\documentclass[letterpaper, 10 pt, conference]{ieeeconf}  
\IEEEoverridecommandlockouts
\usepackage{cite}
\usepackage{amsmath,amssymb,amsfonts}
\usepackage{algorithmic}
\usepackage{graphicx}
\usepackage{textcomp}
\usepackage{xcolor}
\usepackage{hyperref} 
\usepackage{overpic}
\def\BibTeX{{\rm B\kern-.05em{\sc i\kern-.025em b}\kern-.08em
    T\kern-.1667em\lower.7ex\hbox{E}\kern-.125emX}}
\begin{document}

\title{Mixed-Integer Linear Programming Model for Collision Avoidance Planning in Commercial Aircraft Formations\\
\thanks{This research was funded by King Abdullah University of Science and Technology’s baseline support (BAS$/1/1682-01-01$).}
}
\author{Songqiying Yang, Ania Adil, Eric Feron
\thanks{Songqiying Yang, Ania Adil, and Eric Feron are with Computer, Electrical and Mathematical  Science $\&$ Engineering Division (CEMSE), KAUST, Saudi Arabia. (emails:{\tt\footnotesize songqiying.yang@kaust.edu.sa, ania.adil@kaust.edu.sa, eric.feron@kaust.edu.sa} }}

\maketitle

\begin{abstract}
With advancements in technology, commercial aircraft formation flying is becoming increasingly feasible as an efficient and environmentally friendly flight method. However, gaps remain in practical implementation, particularly in collision avoidance for aircraft formations. Existing avoidance algorithms mainly focus on single aircraft or UAV swarms, lacking comprehensive studies on the complex interactions within commercial aircraft formations. To address this, this paper proposes an optimization model designed to generate safe and effective collision avoidance solutions for commercial aircraft formations. This model demonstrates avoidance paths for formations facing intruders and offers insights for developing formation flight strategies. This study explores response strategies for commercial aircraft formations encountering intruders, considering the difficulty of pilot maneuvers. The findings provide theoretical support for the practical implementation of commercial formation flying and may advance the adoption of this technology.
\end{abstract}


\section{INTRODUCTION}
The rapid growth of global air traffic has created an urgent demand for more efficient and environmentally sustainable flight operations. As airlines seek to reduce operating costs and minimize environmental impact, commercial aircraft formation flying has gained increasing attention as a potential solution. The core principle of this technology involves multiple aircraft flying in close formation, where trailing aircraft utilize the wake vortex energy generated by the lead aircraft to reduce aerodynamic drag. By lowering the thrust required to overcome drag, this method significantly reduces fuel consumption and decreases carbon emissions~\cite{b}.

In recent years, numerous flight experiments have further validated the effectiveness and feasibility of formation flying. Experimental data indicate that formation flying can achieve fuel savings of 5\% to 10\% on various flight routes~\cite{b1,b2,b3,b4,b5}. This not only presents significant economic benefits for airlines in terms of fuel costs but also contributes to achieving the global aviation industry's emissions reduction targets. Furthermore, formation flying can be integrated into the existing air traffic management system without requiring major adjustments to current flight routes. Studies have shown that this technology can effectively reduce fuel consumption on both long-haul and short-haul routes, with only minimal adjustments needed to existing flight paths~\cite{b6,b7}.

Despite the obvious advantages of formation flying, its widespread application in commercial aviation is still limited by safety concerns. In military aviation, formation flying has been used for decades, with well-established procedures ensuring the safety of aircraft flying in close proximity. The roles in military formations are predefined, and all maneuvers are planned in advance. Pilots undergo rigorous training and possess extensive experience in executing these operations, ensuring both the safety and efficiency of formation flying. In contrast, the adoption of this technology in commercial aviation has been much slower, primarily due to the unique operational challenges that formation flying presents. First, maintaining precise positioning between commercial aircraft requires extremely high accuracy. Additionally, aircraft must have effective communication systems to ensure timely transmission of information during flight. Most critically, in the event of an emergency, commercial aircraft need to perform urgent maneuvers to ensure safety. However, unlike military formations, the roles of commercial aircraft cannot be predefined, and all maneuvers must be guided by Air Traffic Control, including joining, exiting, and disbanding the formation. These factors make path planning and the practical implementation of formation flying in commercial aviation highly challenging~\cite{b8}.

Currently, research on collision avoidance algorithms for formation flying~\cite{b11,b12} has primarily focused on unmanned aerial vehicles (UAVs), but the maneuverability and operational characteristics of UAVs differ significantly from those of large commercial aircraft. Research on collision avoidance for commercial aircraft~\cite{b13,b14} has focused more on improvements to existing systems, such as the Traffic Collision Avoidance System (TCAS) and Automatic Dependent Surveillance-Broadcast (ADS-B). TCAS provides avoidance instructions when handling potential conflicts between aircraft, while ADS-B helps predict and prevent collisions by transmitting real-time aircraft position information. However, these systems were originally designed to address individual aircraft or small-scale airspace conflicts, and they do not fully account for the complex interactions and dynamic changes involved in managing multiple aircraft formations.

With the continuous advancement of digital communication systems, flight control technology, and autopilot systems, the prospects for the application of commercial formation flying are becoming increasingly feasible. However, despite these technological developments laying the foundation for commercial formation flight, the widespread adoption of this technology still faces a critical challenge—the lack of comprehensive contingency plans for large-scale formations. Previous studies have explored how to generate safe joining, disbanding, and emergency exit paths for aircraft formations at high altitudes~\cite{b9,b10}, which is of great importance in ensuring the basic safety of formations. However, there remains a lack of effective solutions for handling emergencies, such as encounters with intruding aircraft during flight.

To address this issue, this study proposes an optimization model using Mixed-Integer Linear Programming (MILP) to design flight trajectories for aircraft in cruise formation, enabling them to effectively avoid intruders. The goal of this study is to develop a mathematical model that can rival pilot decision-making, helping to devise avoidance strategies when the formation encounters conflict aircraft from different directions. This model ensures that aircraft can safely and swiftly disband the formation to avoid intruders, while meeting the performance constraints of commercial aircraft and avoiding the effects of wake turbulence, before re-forming the formation. {Building upon~\cite{b9,b10}, which explores optimization for generating maneuvering and escape paths across various formation scenarios (such as join, emergency exit, and escape) and evaluates these paths against pilot-generated plans using pilot expertise, this work focuses on designing flight trajectories for aircraft in formation specifically to avoid an intruder aircraft, ensuring safe disbanding and re-formation while considering performance constraints and wake turbulence.} Through this research, we provide new insights into the feasibility of formation flying in commercial aviation and to lay a solid foundation for future research on formation safety and contingency planning.

{In this paper, we first introduce the flight requirements and basic assumptions for the simplified model in Section~\ref{sec_problem} and then translate these requirements into constraints within the MILP model in Section~\ref{sec_MILP}. In Section~\ref{sec_example}, we generate collision avoidance paths for formations consisting of two, three, and five aircraft encountering intruders from the side, and discuss the results. Finally, Section~\ref{sec_conclusion} concludes the paper and proposes directions for future research. 
}

\section{PROBLEM STATEMENT}\label{sec_problem}

\begin{figure}[htbp]
\centering
\begin{overpic}[scale=0.11]{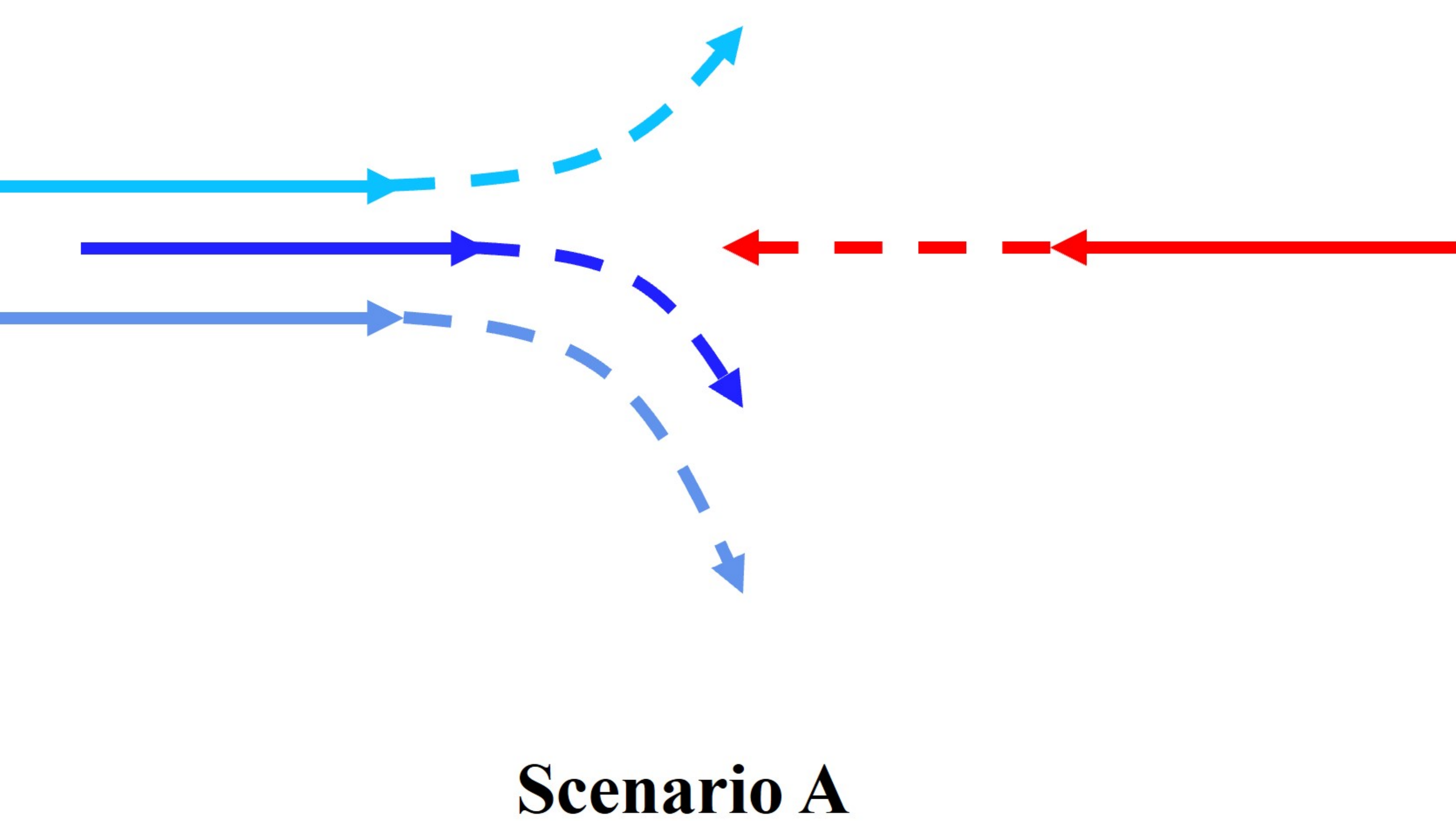}
           \end{overpic} 
\centering
\begin{overpic}[scale=0.11]{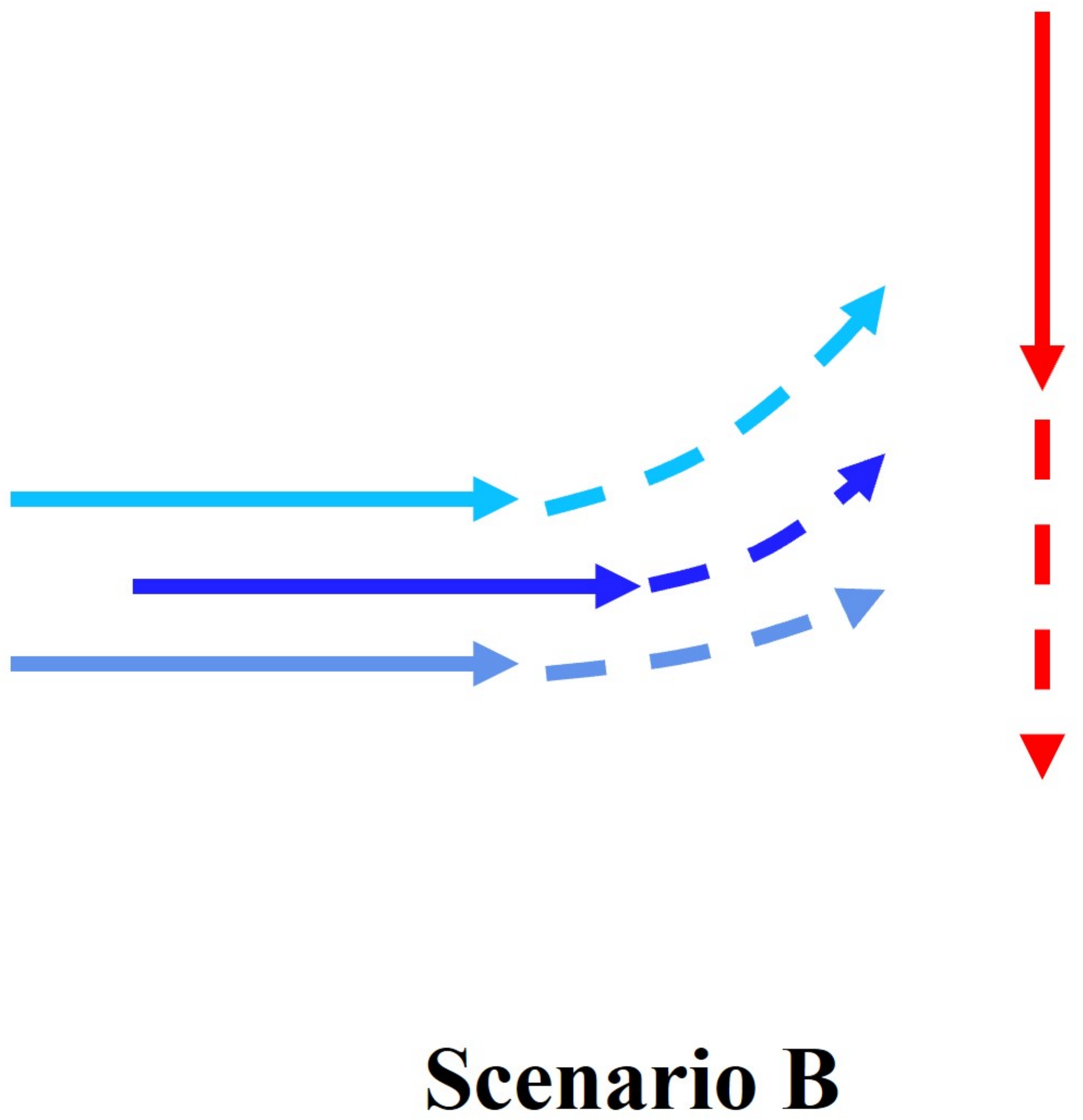}
           \end{overpic} 
\caption{Illustration of possible intruders: scenario A: Head-on Intruder, scenario B: Side Intruder}
\label{fig1-b}
\end{figure}

This model provides theoretical support for generating safe avoidance trajectories for commercial aircraft formations when encountering intruders. In this scenario, the intruder aircraft is usually marked in red, and its flight path intersects with the flight path of the formation. If no maneuver is taken, at least one aircraft in the formation will conflict with the intruder. A conflict is defined as the distance between the two aircraft being less than the specified minimum safe separation. Therefore, to avoid the impending danger, the formation aircraft must maneuver to adjust their flight paths in order to effectively avoid the conflict and ensure safety. Fig.~\ref{fig1-b} illustrates potential intruder scenarios that could be encountered. {The} dashed lines represent the potential maneuvering paths of the formation aircraft in response to intruders approaching from different directions, such as from the front or the side, while the intruder will continue along its original trajectory without any maneuvers.

As commercial aircraft formations have not yet been applied in practice, there are currently no specific standards for commercial aircraft formation. This model draws on formation flight requirements that have been applied in military aviation and incorporates them into the optimization model to ensure that formation operations meet the highest safety standards. For the purposes of modeling and implementation, these requirements have been appropriately simplified, and some model-specific assumptions have been proposed to ensure the model's practicality and effectiveness~\cite{b9,b10}. Addressing these constraints by MILP model will enable the generation of optimal and safe trajectories for commercial aircraft formations, enhancing both safety and operational efficiency in increasingly congested airspaces.

\section{MILP APPROACH}\label{sec_MILP}
{The model operates within a 3D near-inertial reference frame fixed to the Earth at high altitudes, where non-inertial forces are negligible for the short duration of the formation maneuver, which is less than one minute. We define the MILP problem formulation as follows:}
\begin{align}
    \underset{x, v, u}{\text{minimize}}  = &J_{\text{maneuver}}(x, v, u) + J_{\text{avoidance}}(x)\notag \\
    &+ J_{\text{drag}}(x) + J_{\text{smoothness}}(x)
\end{align}
\begin{align}
    \text{subject to},\notag\\
    &\forall i \in \{1, \ldots, T\},\notag \\
    &\forall (p, q) \in \{1, \ldots, A\}^2, q > p,\notag \\
    &\forall r \in \{1, \ldots, NI\}, \forall d \in \{1, \ldots, 3\}, \notag \\
    &x_{(i+1)pd} = x_{ipd} + (v_{ipd} + W_d) \Delta t, \notag \\
    &v_{(i+1)pd} = v_{ipd} + u_{ipd} \Delta t, \notag \\
    &v_{ipd} \in \mathcal{V}, \quad u_{ipd} \in \mathcal{U}, \notag \\
    &x_{ipd} \notin \mathcal{S}_q, \quad x_{ipd} \notin \mathcal{W}_q,\notag \\
    &x_{ipd} \notin \mathcal{S}_r, \quad x_{ipd} \notin \mathcal{W}_r
\end{align}

The aircraft model is formulated using a simple discrete-time system. The cost function includes the maneuvering cost ($J_{\text{maneuver}}$), the time cost required to avoid intruding aircraft ($J_{\text{avoidance}}$), the induced drag cost due to the increased separation of formation aircraft ($J_{\text{drag}}$), and the smoothness cost of the avoidance p[ath ($J_{\text{smoothness}}$).  {The MILP model is based on the formulation proposed in~\cite{b9,b10}. Due to space constraints, only the new constraints, 
$J_{\text{drag}}$ and $J_{\text{smoothness}}$, will be presented here, with the remaining details available in~\cite{b9,b10}.} The design variables—position, velocity, and acceleration are represented by \( x_{ipd} \), \( v_{ipd} \), and \( u_{ipd} \), respectively, where \( i \) denotes the time step, \( p \) denotes the aircraft, and \( d \) denotes the spatial dimension. 

Note that \( x_{ird} \), \( v_{ird} \), and \( u_{ird} \) are not design variables, since the velocity of the intruder is constant, and its acceleration is set to zero. The intruder's position at each time step is predetermined as \( x_{(i+1)rd} = v_{ird} \Delta t \). The wind effect is incorporated into the model through the variable \( W_d \); however, for the sake of simplicity, this paper considers only calm wind scenarios, setting \( W_d = 0 \). The aircraft's velocity is constrained within the permissible range \( \mathcal{V} \), and acceleration is bounded by \( \mathcal{U} \). The subscripts \( p \) and \( q \) are used to denote different aircraft, \(r\) is used to denote the intruder, with the position of aircraft \( p \) at time \( i \) in dimension \( d \), \( x_{ipd} \), required to maintain a minimum separation from other aircraft \( q \), defined by \( \mathcal{S}_q \), a minimum safe distance from the intruder, defined by \(\mathcal{S}_r\), and to avoid entering the wake turbulence zone of any other aircraft \( q \) and \(r\), represented by \( \mathcal{W}_q \) and \(\mathcal{W}_r\).
\subsection{Cost Functions}
During the cruise, aircraft achieves maximum lift-to-drag ratio with balanced forces. However, thrust demand significantly increases during acceleration or climbing, so the model should minimize maneuvering actions wherever possible in order to reduce fuel consumption. To mitigate the risk of losing visual contact during obstacle avoidance, the formation should dissolve and re-establish as quickly as possible, ensuring that the avoidance process is completed in the shortest feasible time. The time cost  penalizes any aircraft that are not at their designated positions in the course at each time step. {The cost functions $J_{\text{maneuver}}$ and $J_{\text{avoidance}}$ can be found in~\cite{b9}. }

The degree to which the leading aircraft reduces induced drag for the trailing aircraft is highly dependent on their relative position. We define the position that maximizes this reduction as the optimal position relative to the nearest preceding aircraft. The induced drag cost, \(J_{\text{drag}}\), penalizes deviations of the trailing aircraft from the optimal position in any axis.

\begin{align}
&J_{\text{drag}} = \sum_{i=1}^{T} \sum_{p=2}^{A} \sum_{d=1}^{3} \omega_h h_{ipd}, \notag \\
&\forall i \in \{1, \ldots, T\}, \forall p \in \{2, \ldots, A\}, \notag \\
&h_{ipd} \geq |x_{ipd} - H_{ipd}| \quad \text{with} \quad \forall d \in \{1, \ldots, 3\}
\end{align}
where \(h_{ipd}\) is the drag cost slack variable in \(ft\) and \(H_{ipd}\) is the optimal position. Within the formation, drag can be reduced by up to approximately \(33\%\) compared to outside the formation, with reductions observed when deviating from the optimal position~\cite{b7}. Therefore, the impact on thrust from deviating from the optimal position is much smaller than that from aircraft maneuvers. As we set the maneuver cost function weight \(\omega_g\) to 1 \(s^2/ft\), the drag cost function weight \(\omega_h\) is set to be 0.25 \(ft^{-1}\).

To facilitate manual control, the solution needs to reduce the frequency of vertical maneuvers and avoid situations where the aircraft ascends, descends, and then ascends again in a short period. The smoothness cost \(J_{\text{smoothness}}\) penalizes excessive variations in the aircraft's vertical trajectory.

\begin{align}
&J_{\text{smoothness }} = \sum_{i=2}^{T} \sum_{p=1}^{A} \omega_k k_{ipd}, \notag \\
&\forall i \in \{2, \ldots, T\}, \quad \forall p \in \{1, \ldots, A\},   \notag \\
&k_{ipd} \geq |x_{ipd} - x_{(i-1)pd}|,   d = 3
\end{align}
where \(k_{ipd}\) is the smoothness cost slack variable in ft and the smoothness cost weight \(\omega_k\) needs to be adjusted to balance the trade-off between smoothness and maneuvering cost, with different values of \(\omega_k\) used depending on the scenario. The specific choices of \(\omega_k\) will be discussed in detail in IV.

\subsection{Performance Constraints}
This model is based on a performance limitations of narrow-body aircraft during high-altitude flight, as in ~\cite[Table 1]{b10}.

\subsection{Avoidance Constraints}

{The model incorporates three avoidance constraints. As discussed in Section~\ref{sec_problem}, each aircraft must avoid collisions with other aircraft within the formation. A minimum safety distance (\(\mathcal{S}_{q}\)) is enforced between all aircraft in the formation. Moreover, to prevent interference from wake turbulence, all aircraft must steer clear of the wake turbulence generated by any other aircraft in the formation~\cite{b9}. Furthermore, a new constraint is introduced to account for intruding aircraft. This constraint ensures that any intruder aircraft maintains a safe distance from the formation, preventing potential collisions. The minimum safe distance for the intruder $\mathcal{S}_{r}$ is defined below. 
}

\subsubsection*{Intruder Minimum Safe Distance (\(\mathcal{S}_{r}\))} 
The intruder aircraft can be of any type, but here we use a narrow-body aircraft as an example. If a different aircraft type is used or if higher safety requirements are needed, the value of \( R_d \) can be adjusted accordingly. The constraint is expressed
\begin{align}
&\forall i \in \{1, \ldots, T\},  \notag \\
&\forall p \in \{1, \ldots, A\},  \forall r \in \{1, \ldots, NI\}, \notag \\
&x_{ipd} - x_{ird} \geq R_s - M a_{iprds} \notag \\
&\text{with} \quad s = 1, \quad \forall d \in \{1, \ldots, 3\}, \notag \\
&x_{iqd} - x_{ird} \geq R_s - M a_{iprds} \notag \\
&\text{with} \quad s = 2, \quad \forall d \in \{1, \ldots, 3\}, \notag \\
&\sum_{d=1}^{3} \sum_{s=1}^{2} a_{iprds} \leq 5
\end{align}
where \(a_{iprds}\) is a binary variable. At each time index \(i\), the position of each aircraft must maintain a \(R_{s} \) separation from all intruders. For intruders, the safe distance(\(R_{s}\)) is set at 1500ft in \(d = 1\) and \(d = 2\), 600ft in \(d = 3\).

\subsection{Optimization}
The model utilizes the commercial Gurobi optimization software, implemented in Python on the high-performance computing cluster Ibex at KAUST. To achieve finer detail, the typical time step \( \Delta t \) in the model is set between 0.8 and 1.2 seconds. For longer time steps, due to the discrete nature of the model, some collisions may go undetected. Specifically, when the time step is large, rapid approaches or crossings between aircraft may occur between time steps, and these details may not be captured, leading to undetected conflicts.

For scenarios involving two aircraft, the model finds a solution in 30 minutes, attaining a relative optimality gap of less than $1\%$ in the mixed-integer programming (MIP) solution. When the scenario involves three aircraft, the model requires several hours to generate a solution. As the number of aircraft increases to five or more, the computation time extends to one or two days, with a relative MIP gap of approximately 5\%.

\section{ILLUSTRATIVE EXAMPLE}\label{sec_example}
As a test case for our MILP model, the scenario A where the intruder approaches from the side at a 90-degree angle, as shown in Fig.~\ref{fig1-b}, is considered. This scenarios is useful as an intruder approaching the formation from any direction can be viewed as a combination of this situations. 

\begin{figure}[htbp]
\centering
\includegraphics[width=0.5 \linewidth]{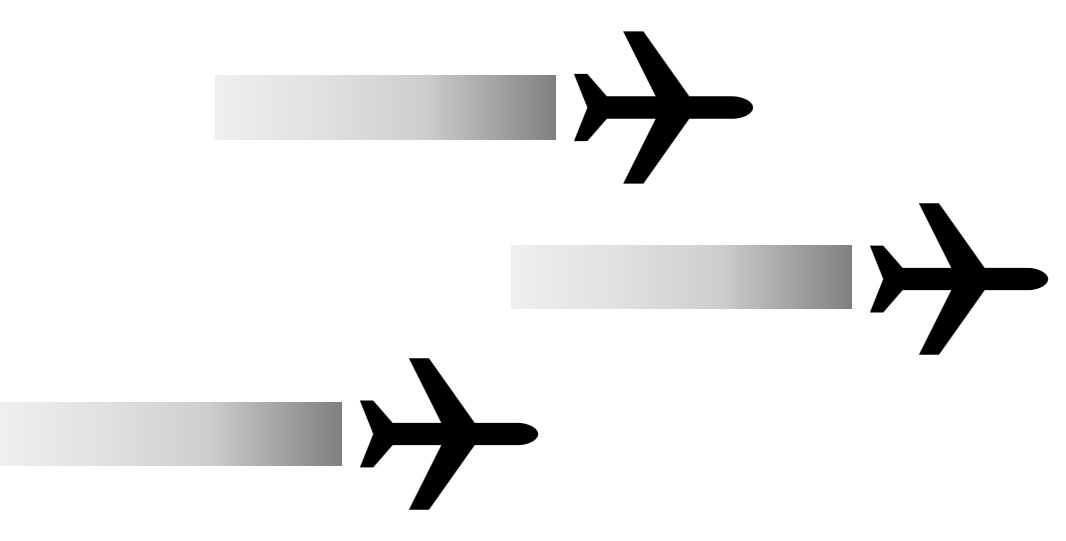} 
\caption{Illustration of the initial formation setting.}
\label{fig2}
\end{figure}

At the initial state, the formation advances in a V-shape (as in Fig.~\ref{fig2}) along the course (positive \(x_1\) direction) at a speed of 750 ft/s, with no lateral or vertical speed and acceleration. The intruder is at the same altitude as the formation and is positioned along the centerline of the course (on the $x_1$ axis), 35,000 ft away from the formation's initial position. The intruder is positioned 15,000 ft to the right of the formation's course and 15,000 ft away from the formation's starting position. The intruder moves laterally toward the formation at a constant speed of 750 ft/s, the same as the formation's speed. The intruder and the formation are at the same altitude. If the formation takes no evasive maneuvers, a collision is expected to occur at 10 seconds.
We stipulate that during the final five time intervals, the formation aircraft must have no lateral or vertical velocity or acceleration, return to the original course boundaries, and continue traveling along the course at the initial speed.
\subsection{The 2-aircraft Avoidance} 
\begin{figure}[htbp]
\centering
\begin{overpic}[scale=0.11]{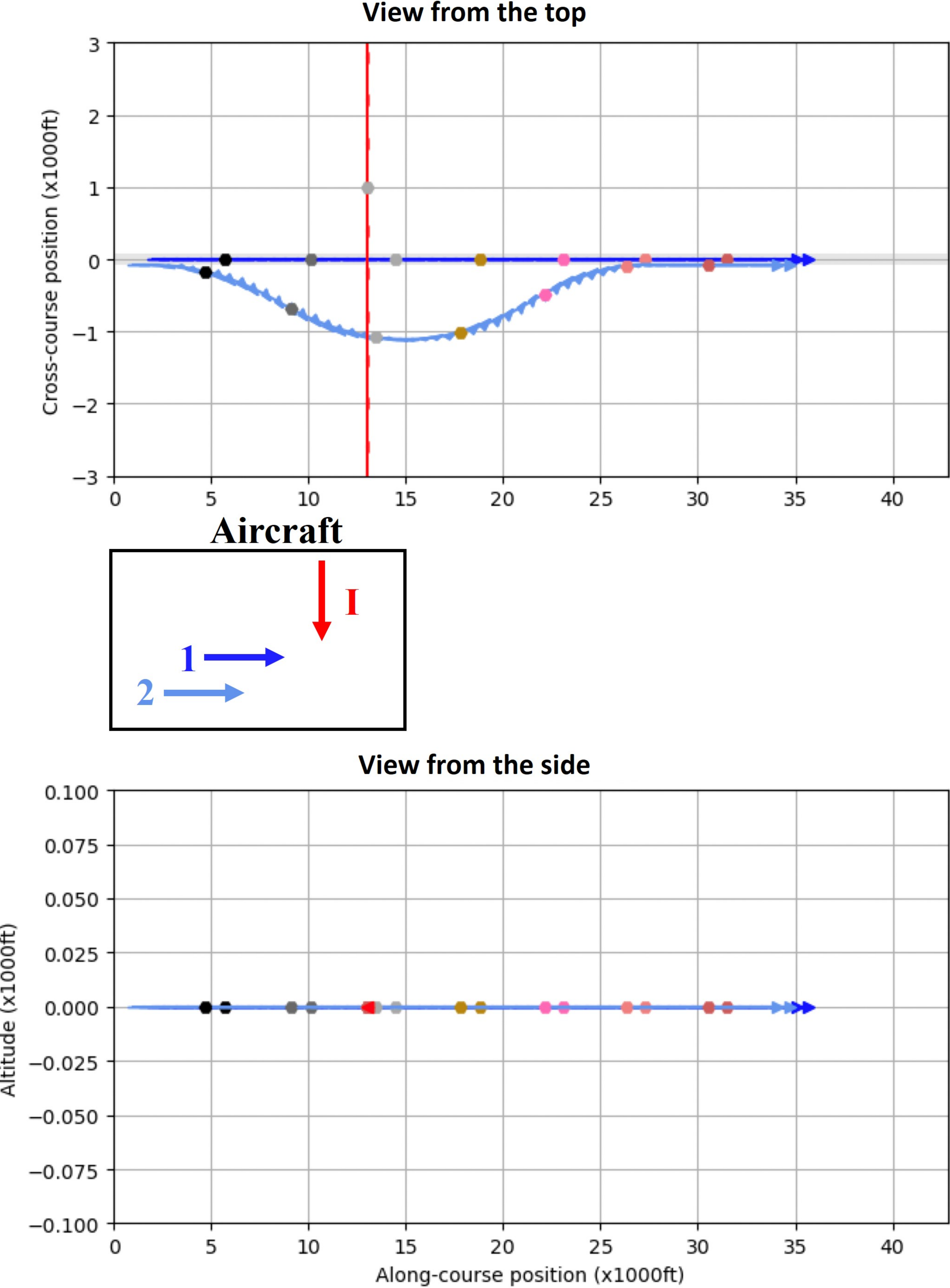}
           \end{overpic} 
\caption{2-aircraft avoidance solution of side intruder.}
\label{fig9}
\end{figure}

\begin{figure}[htbp]
\centering
\begin{overpic}[scale=0.2]{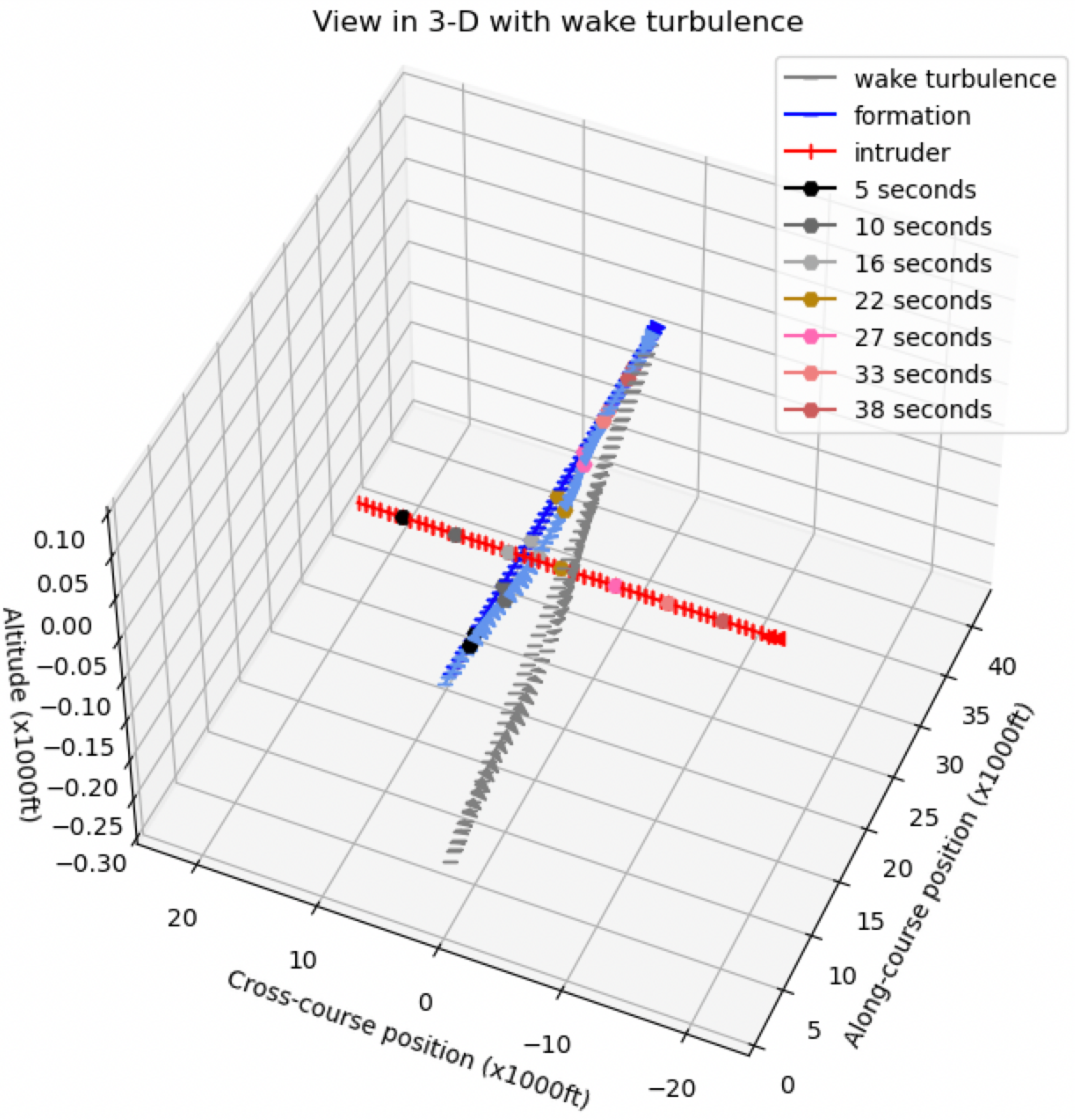}
           \end{overpic} 
\caption{3D View of 2-aircraft avoidance solution of side intruder.}
\label{fig10}
\end{figure}

Fig.~\ref{fig9} illustrates the avoidance paths of a two-aircraft formation when encountering a lateral intruder. To make the lateral maneuvers more visible, the vertical axis of the top-down view has been magnified, which prevents the full display of the intruder’s trajectory. However, the complete approach of the intruder is shown in Fig.~\ref{fig10}. In the side view, since the intruder is moving at the same altitude as the formation from the side, its trajectory is projected as a single point.

It can be observed that Aircraft 1 continues along its course, indicating that there is no risk of collision with the intruder. In contrast, Aircraft 2 performs a lateral maneuver to the left (away from the intruder) before returning to the center of the course. This avoidance strategy relies entirely on lateral maneuvering, with all aircraft maintaining a constant altitude during the avoidance process, making the maneuver relatively easy for pilots to execute.

\subsection{The 3-aircraft Avoidance} 
\begin{figure}[htbp]
\centering
\begin{overpic}[scale=0.11]{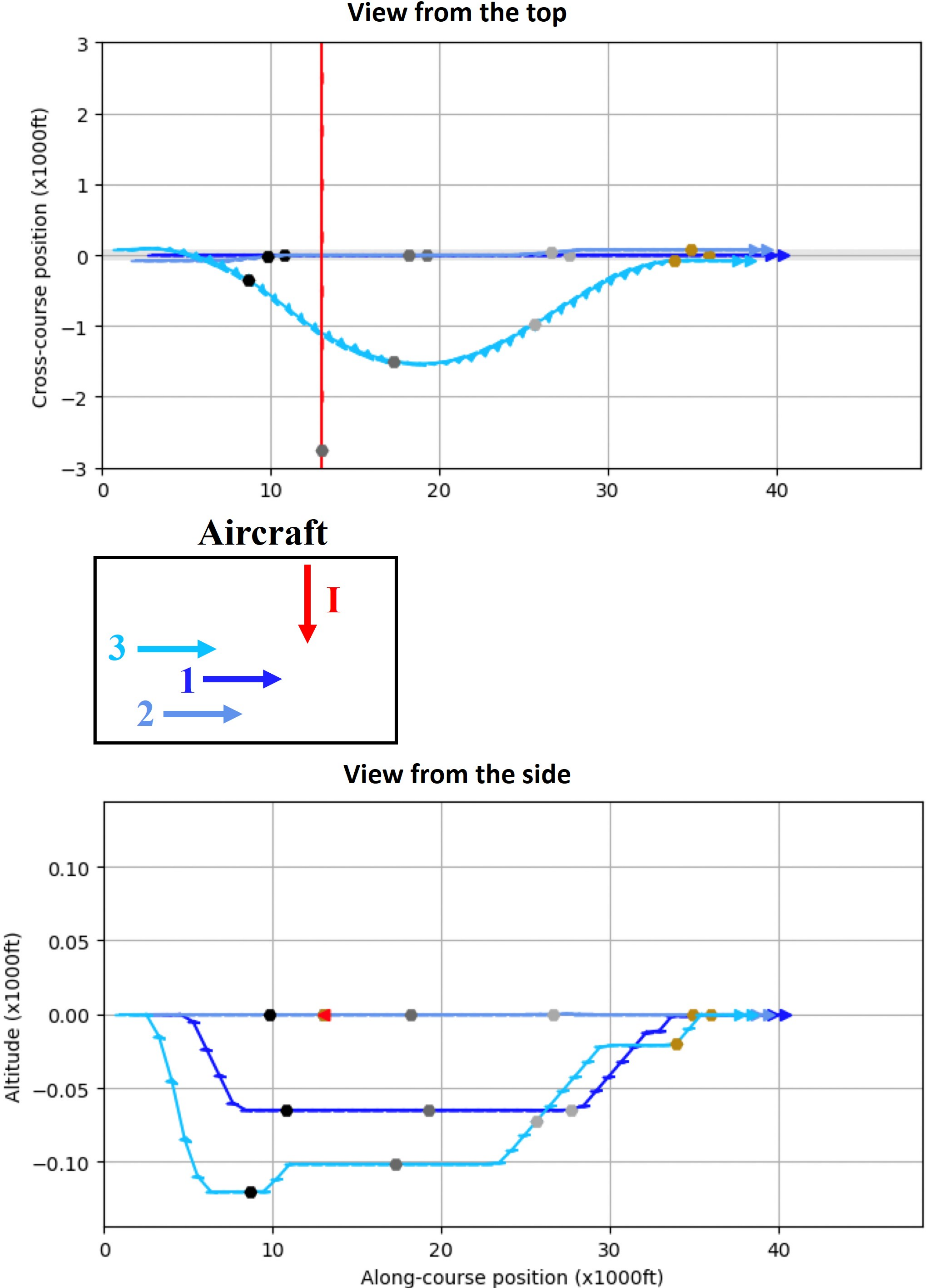}
           \end{overpic} 
\caption{3-aircraft avoidance solution of side intruder.}
\label{fig11}
\end{figure}

Fig.~\ref{fig11} presents the avoidance paths of a three-aircraft formation when encountering a lateral intruder. Compared to the two-aircraft formation, the three-aircraft avoidance paths are more complex and include vertical maneuvers. Aircraft 1 maintains a stable lateral trajectory without any lateral maneuvering. Aircraft 2 maneuvers from the right side of Aircraft 1 to the left to avoid the intruder, while Aircraft 3 maneuvers to the right, tracing a larger arc to avoid the intruder. In the vertical direction, Aircraft 1 and Aircraft 3 descend to a lower altitude and later return to their original altitude, while Aircraft 2 does not perform any vertical maneuvering.

From the two-aircraft formation scenario, the intruder does not pose a threat to Aircraft 1. However, in the three-aircraft solution, Aircraft 1 descends as Aircraft 2 maneuvers to the right, mainly to avoid the wake turbulence generated by Aircraft 2. Unlike in the two-aircraft scenario, Aircraft 2 does not make a large rightward arc, but instead makes a slight leftward adjustment. This modification aims to achieve the overall optimal path for the formation. Additionally, it can be observed that in the final formation recovery, Aircraft 2 and Aircraft 3 switch positions. This swapping of positions enhances the flexibility of the aircraft, allowing them to better avoid the intruder.

We can observe that the position of Aircraft 1, the lead aircraft, remains unchanged after the maneuver. This is due to the fact that, in formation flying, the lead aircraft not only guides the direction but also plays a crucial role in commanding and coordinating the entire formation. The lead aircraft is indispensable in ensuring that the formation maintains its course, speed, and altitude. Moreover, air traffic control (ATC) coordinates with the lead aircraft as the reference point, and the lead aircraft’s heading and speed directly influence the safety and efficiency of the entire formation. Therefore, regardless of the avoidance strategy, the position and role of the lead aircraft cannot change to ensure the overall stability and safety of the formation.

\begin{figure}[htbp]
\centering
\begin{overpic}[scale=0.2]{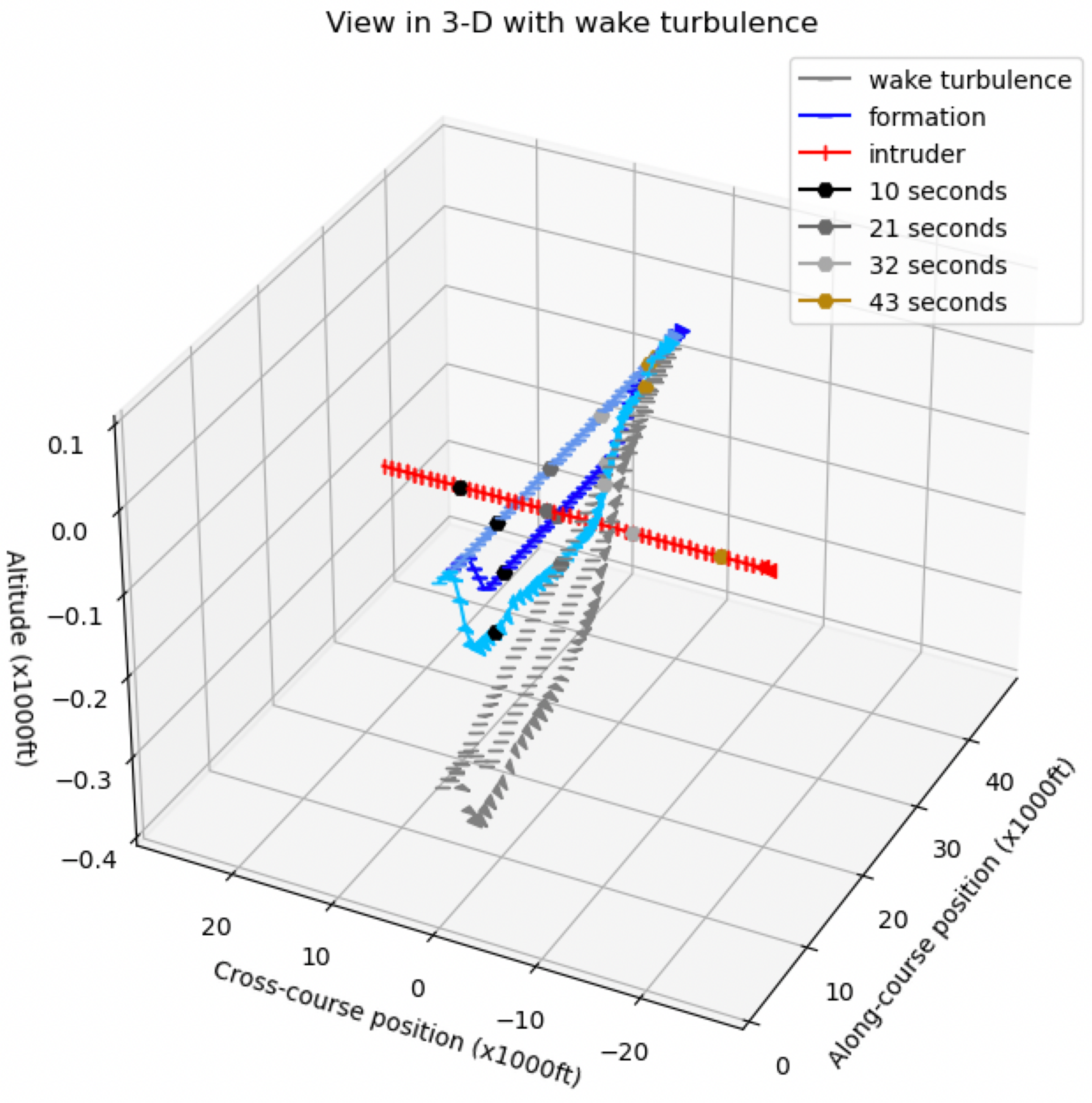}
           \end{overpic} 
\caption{3D View of 3-aircraft avoidance solution of side intruder.}
\label{fig12}
\end{figure}
\subsection{The 5-aircraft Avoidance} 
\begin{figure}[htbp]
\centering
\begin{overpic}[scale=0.11]{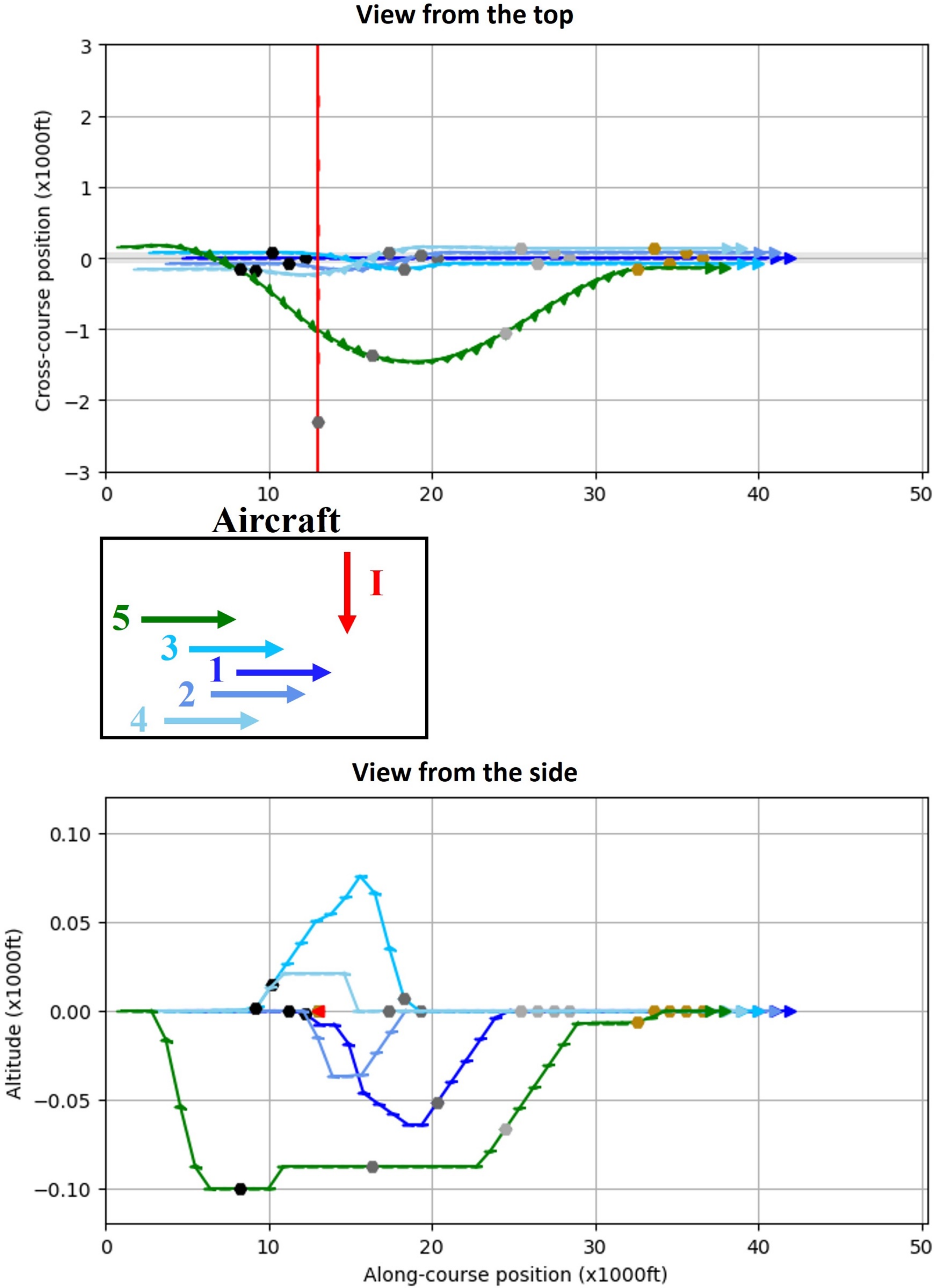}
           \end{overpic} 
\caption{5-aircraft avoidance solution of side intruder.}
\label{fig13}
\end{figure}

Fig.~\ref{fig13} illustrates the avoidance paths of a five-aircraft formation when encountering a lateral intruder. Compared to the three-aircraft formation, the avoidance strategy for the five-aircraft formation is more complex, involving multiple maneuvers in both lateral and vertical directions. The five-aircraft formation requires a longer time (\(T = 50s\)) to complete the avoidance. {To achieve simpler results, we tested outcomes with weights ranging from 0 to 20, in intervals of 5. A smoothness weight of \(w_k = 10\) proves to be a better choice, as each aircraft only needs to perform a single vertical maneuver.} {This choice significantly reduces the computational cost, as fewer maneuvers result in fewer decision variables and constraints within the optimization problem. Moreover, it simplifies trajectory adjustments, making the solution easier to interpret and more practical to implement.}

As shown in the figure, the lead aircraft (Aircraft 1) undergoes the least overall maneuvering, performing only one downward vertical maneuver without any lateral adjustments. In contrast, Aircraft 5 executes the most extensive maneuvers, including a wide rightward arc and a descent to a lower altitude to avoid the intruder and the wake turbulence from other aircraft. In the restructured formation, Aircraft 2 and Aircraft 4 have shifted from the right side of the lead aircraft to the left, while Aircraft 3 and Aircraft 5 have moved in the opposite direction.

It can be inferred that for formations consisting of three or more aircraft, swapping the positions of the aircraft on either side of the lead aircraft after avoiding a conflict is a favorable strategy to mitigate further conflicts. Additionally, it is observed that when facing a lateral intruder, the aircraft on the outermost side of the formation, which is closest to the intruder, requires the most significant maneuvering. However, the degree of this maneuver is considerably smaller. This situation also depends on the distance between the formation and the intruder at the point of detection, which warrants further investigation.

\begin{figure}[htbp]
\centering
\begin{overpic}[scale=0.2]{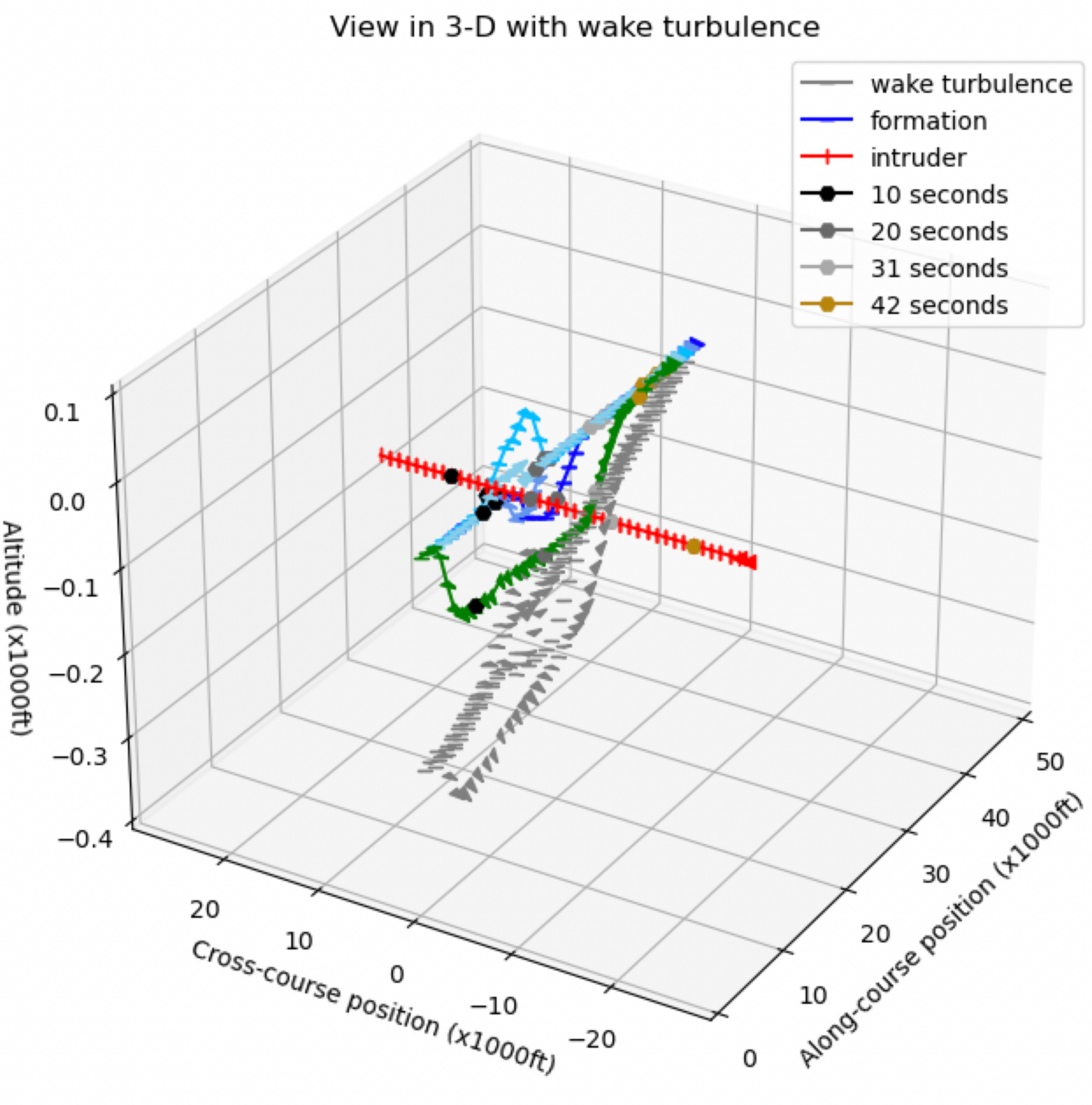}
           \end{overpic}
\caption{3D View of 5-aircraft avoidance solution of side intruder.}
\label{fig14}
\end{figure}

In conclusion, Fig.~\ref{fig13}, Fig.~\ref{fig14} demonstrates that the five-aircraft formation successfully avoids the intruder through flexible lateral and vertical maneuvers, while maintaining adequate safety distances during the avoidance process.

\section{CONCLUSIONS}\label{sec_conclusion}
This study proposes an optimization model for generating collision avoidance solutions for commercial aircraft formations, providing valuable insights for air traffic control and pilots. The model employs an MILP approach and successfully generates safe and operable flight paths while satisfying constraints related to aircraft performance, formation flight requirements, and wake turbulence avoidance. The study chooses one important scenario to generate and evaluate the solutions and derives several insightful conclusions for designing avoidance paths: 1. A formation consisting of two aircraft can successfully avoid intruders solely through lateral maneuvers, whereas formations with more aircraft must introduce vertical maneuvers to effectively avoid intruders. 2. For formations with three or more aircraft, swapping the positions of the aircraft on either side of the lead aircraft after avoiding a collision is a favorable strategy. 3. For lateral intruders, the aircraft on the outermost side of the formation, closest to the intruder, requires larger maneuvers, which also depends on the distance between the formation and the intruder at the point of initial contact.

Future research will test the model in more scenarios and focus on developing more robust solutions by identifying a reasonable range of intruder positions, such that when an intruder falls within this range, the formation can avoid it by adopting the same maneuvering path. This approach aims to enhance the model’s applicability and reliability in practical operations.

\bibliography{biblio}

\end{document}